\documentclass[runningheads]{llncs}
\usepackage{marvosym}
\usepackage{verbatim}
\usepackage{listings}
\usepackage{courier}
\usepackage{isabelle}
\usepackage{isabellesym}
\usepackage{graphicx}
\usepackage{amsmath,amssymb,amsfonts}
\usepackage{multirow}
\usepackage{tikz}

\renewcommand{\implies}{\longrightarrow}

\newcommand{\holpy}{\textsf{HolPy}}

\begin{document}
\title{\holpy: Interactive Theorem Proving in Python (System Description)}
\author{Bohua Zhan$^{\textrm{(\Letter)}}$}
\authorrunning{B. Zhan}
\institute{State Key Laboratory of Computer Science, Institute of Software, \\
    Chinese Academy of Sciences, Beijing, China \\
  \email{bzhan@ios.ac.cn}}
\maketitle              % typeset the header of the contribution

\begin{abstract}
  \holpy\ is an interactive theorem proving system implemented in
  Python. It uses higher-order logic as the logical foundation.
  Its main features include a pervasive use of macros in producing,
  checking, and storing proofs, a JSON-based format for theories, and
  an API for implementing proof automation and other extensions in
  Python. A point-and-click-based user interface is implemented for
  general-purpose theorem proving. We describe the main design
  decisions of \holpy, current applications, and plans for the future.

\end{abstract}

\section{Introduction}

\holpy\footnote{Available at \url{https://gitee.com/bhzhan/holpy}}
is a new interactive theorem proving system implemented in Python.
Like Isabelle/HOL \cite{isabelle}, HOL Light \cite{hol-light}, and HOL4 \cite{hol4},
it is based on higher-order logic (simple type theory).
As such, the aim of \holpy\ is not to propose any innovation in the logical
foundations, but rather in the architecture of the system, in particular the representation
of proofs and theories, and the approach to proof automation and other extensions.

There are three features that lie at the core of \holpy's design:
\begin{enumerate}
\item Pervasive use of macros in producing, checking, and storing proofs.
The concept of macros is also present in systems such Lean \cite{lean}. We demonstrate
that it can be usefully applied in a system based on higher-order logic.
It makes generating explicit proof terms scalable to large proofs,
in particular those involving a large amount of computation
(Section \ref{sec:proof-representation-macros}).

\item A JSON-based format for theories. JSON is a widely-used format for storing
and exchanging structured data. Data in JSON format can be naturally considered as objects in both
Python and JavaScript. Compared to using a text-based format for theories,
this design may make implementing an user interface initially more difficult.
However, in the long run it allows greater flexibility in its design (Section \ref{sec:theory-format}).

\item A Python-based API for implementing proof automation and other extensions
to the system. From the API a programmer can access all aspects of the system, including
construction of terms and proof terms, and definition of new macros
(Section \ref{sec:python-api}).

\end{enumerate}

The intention of these design choices is twofold. First, by exposing a Python-based API
for proof automation and other extensions, we aim to make extension of the system
accessible to a wider user base. Second, the JSON-based theory format
means it is possible to implement different user interfaces for different domains
of application, but nevertheless produce theories that are compatible with each other.

In the next three sections, we explain the three main design features in more detail.
In Section \ref{sec:user-interface}, we briefly discuss one general-purpose
and one specialized user interface for \holpy, the details of which are left to
other papers. Finally, we review related work in Section \ref{sec:related-work} and conclude in
Section \ref{sec:conclusion}.

\section{Proof representation with macros}\label{sec:proof-representation-macros}

\subsection{Proof rules and macros}\label{sec:proof-rules-macros}

Proofs in \holpy\ are conducted in natural deduction style. A sequent
(class \textsf{Thm} in Python) has the form $A_1,\dots,A_n \vdash C$, stating
that the (unordered) list of antecedents $A_1,\dots,A_n$ imply the consequent $C$.
A \emph{proof rule} is a function which takes a list of input sequents and
outputs a sequent which logically follow from the inputs. The evaluation of
a proof rule may depend on the background theory, and each proof rule may take
additional arguments depending on the rule.

There are two classes of proof rules: primitive derivation rules and macros.
The primitive derivation rules correspond to the basic rules of reasoning in
higher-order logic, such as introduction and elimination of implies and
forall quantification, substitution rules, transitive and congruence rules of equality,
and so on. Special proof rules include \textsf{theorem}, which takes as argument
the name of a theorem, and attempts to look up that theorem in the current
theory environment. The primitive rule \textsf{sorry} takes a sequent as argument
and simply outputs that sequent. It represents a gap in the current proof.

The concept of macros is similar to that in \cite{lean}. Macros can be considered
abbreviations of several applications of more basic proof rules. In addition to the
evaluation function, a macro may additionally implement an expansion
function. When invoked, it returns a proof (in the format defined below) for
the output sequent. The expanded proof can be used during proof checking, so the
evaluation of the macro does not have to be trusted.

The use of macros greatly reduces the size of proofs that need to be stored. It
can also greatly reduce the time needed to check a proof (if certain macros are
trusted). To give a simple example: evaluations of arithmetic operations on
natural numbers eventually reduce to applying theorems about binary representations
of numbers. This can become slow as the numbers involved get large. The corresponding
proof term, if generated, would also take up a large amount of memory. Instead, it
is possible to define evaluation of arithmetic operations as a macro. Its
evaluation simply uses Python's native operations on numbers. If the
evaluation needs to be checked, the expansion function is then used to generate
on demand the proof based on binary representations. Further savings of both
time and memory can be achieved, for example, by implementing normalization of
polynomials on natural numbers as a macro.

\subsection{Proof representation}\label{sec:proof-representation}

Proofs are represented in two ways in \holpy: as proof terms and as linear
proofs. Proof terms represent proofs as rooted directed acyclic graphs
in memory. Each vertex of the graph corresponds to application of one proof
rule, using references to point to input sequents. Proof terms are easy to
construct using Python code, and hence are convenient for proof automation.

Once a proof term is constructed, it can be converted to linear proofs for checking,
storage, and display to the user. A \emph{linear proof} is an ordered list of
proof items. Each proof item consists of an identifier, the name of a proof rule,
a list of identifiers of earlier proof items in the list (referring to the
input sequents), and additional arguments of the rule. By convention,
identifiers are tuples of natural numbers in dot-separated form (e.g.
$n_1.n_2\dots n_k$). This allows expressing subproofs (e.g. the steps with
identifiers $1.0, 1.1, 1.2$, etc. form the subproof for step 1).

\section{JSON-based theory format}\label{sec:theory-format}

Theories are stored in the JSON format. JSON (for JavaScript Object Notation \cite{json})
is a standard for data exchange designed for the web. It stores
data as a hierarchy of dictionaries and lists, with strings, numbers and booleans
as possible atomic values. Data in JSON format can be easily translated to
objects in both Python and JavaScript. Hence, conceptually we can consider
a theory itself as a Python object (specifically a dictionary) that can be
generated and manipulated just like any others.

\subsection{Format of a theory}

At the top level, a theory is a dictionary with keys \textsf{name} for
name of the theory, \textsf{imports} for the list of imported theories,
and \textsf{content} for the main content. The content is a list of
items. Each item can be an axiomatic constant, a definition, an axiom,
a theorem, an inductive datatype, and so on. Each item is a dictionary
with key \textsf{ty} specifying the type of the item. The interpretation
of other fields depends on the type of the item. We use theorems
as an example. A theorem item contains keys \textsf{name} for name of
the theorem, \textsf{vars} for its free variables (dictionary mapping names
of variables to its type), \textsf{prop} for the statement of the
theorem as a string, and \textsf{proof}, containing the proof of the theorem.
The proof is represented as a linear proof, following the format given in
Section \ref{sec:proof-representation}. A theorem item can optionally contain
other keys, such as attributes (analogous to Isabelle's attributes,
useful for search tools in the user interface as well as proof automation).

Compared to a text-based theory format, this format makes user interfaces more
difficult to implement at first. However, we believe it offers significant
advantages in the long run. For example, information that does not need to be seen
by the user but are important for other purposes (e.g. certificates generated
by external proof automation) can be stored in the file.

\subsection{Checking a theory}

How to guarantee correctness of proofs based on a small trusted kernel is
very important for an interactive theorem prover. The initial choice of
ML-family of languages is to a large extent due to its high type and memory safety guarantees.
Traditional imperative languages such as C++, Java, and Python offer
no such guarantees. We approach this problem by allowing theories to be checked
by external tools. Indeed, a theory in JSON format contains all information needed
to check its correctness. First, statements of earlier theorems are loaded from imported theories.
Then, the proofs of theorems are checked one-by-one. If a macro needs to be expanded,
the external tool can either use its own implementation of the expansion function
(for particularly common macros) or invoke the reference Python implementation to
obtain its expansion. In this way, proofs can be checked independently by
multiple tools written in different languages, offering a very high level of
guarantee for its correctness.

\section{Python-based API}\label{sec:python-api}

In this section, we describe the Python-based API for implementing proof automation
and other extensions. The aim is to show that the API for proof automation that is
traditionally based on functional programming can be simulated in an imperative
language like Python.

\subsection{Terms, theorems, and proofs}

At the lowest level, the classes \textsf{HOLType}, \textsf{Term}, and \textsf{Thm}
represent higher-order logic types, terms, and sequents, respectively.
The class \textsf{ProofTerm} represents a proof term (a vertex in
the directed acyclic graph). The \textsf{th} field of a proof term is the sequent
proved. The class offers functions for constructing new proof
terms from old ones. For example:
\begin{itemize}
\item \textsf{ProofTerm.implies\_elim}($pt_1,pt_2$): with $pt_1$ of form $A\implies B$
and $pt_2$ of form $A$, get proof term for $B$.
\item \textsf{ProofTerm.subst\_type}($\mathit{tyinst},pt$): substitute type
instantiation $\mathit{tyinst}$ into $pt$.
\item \textsf{ProofTerm.substitution}($\mathit{inst},pt$): substitute term
instantiation $\mathit{inst}$ into $pt$.
\end{itemize}
These functions can be combined to form more complex procedures for constructing proof terms.

\subsection{Macros}

Macros (defined by constructing proof terms in the expansion) are
represented as class \textsf{ProofTermMacro}, and new macros are defined by inheriting
from this class. Parameters of the macro can be set in its \textsf{\_\_init\_\_}
method. This includes its trust level and signature. The \textsf{eval} function computes the output
sequent from a list of input sequents (\textsf{Thm} objects). The \textsf{get\_proof\_term}
function computes the expanded proof (as a proof term) from a list of input proof terms.

As an example, we show a simplified version of the macro for applying an existing theorem in the theory:
\begin{lstlisting}[language=Python]
class apply_theorem(ProofTermMacro):
   def __init__(self):
      self.level = 1
      self.sig = str

   def eval(self, thy, name, prevs):
      th = thy.get_theorem(name, svar=True)
      ts = [prev_th.prop for prev_th in prevs]
      instsp = matcher.first_order_match_list(th.assums, ts)
      C = subst_norm(th.concl, instsp)
      return Thm(sum([prev.hyps for prev in prevs], ()), C)

   def get_proof_term(self, thy, name, pts):
      th = thy.get_theorem(name, svar=True)
      ts = [prev_th.prop for prev_th in prevs]
      tyinst, inst = matcher.first_order_match_list(th.assums, ts)
      pt = ProofTerm.theorem(thy, name)
      pt = ProofTerm.substitution(tyinst, ProofTerm.subst_type(inst, pt))
      for prev_pt in pts:
         pt = ProofTerm.implies_elim(pt, prev_pt)
      return pt
\end{lstlisting}
In the \textsf{\_\_init\_\_} function, the trust level is set to 1 (most trustworthy)
and the signature is set to string (taking name of a theorem as argument). In the
\textsf{eval} function, the theorem with the given name is loaded from the current
theory environment. The function then matches the assumptions with the propositions
of input theorems, and returns the instantiated conclusion.
In the \textsf{get\_proof\_term} function, the expanded proof in terms of \textsf{theorem},
\textsf{substitution}, \textsf{subst\_type} and \textsf{implies\_elim} rules are
constructed. With this macro defined (and registered to the global list of macros),
proof terms can simply invoke the \textsf{apply\_theorem} rule to apply
an existing theorem in one step.

\subsection{Conversions and tactics}

The concept of conversions is essential to automation of rewriting in the ML-family of languages.
A conversion is a function taking a term $t$ as input and returns a theorem of
the form $t=t'$. In our API, the class \textsf{Conv} defines the concept of
conversions, and new conversions are defined by inheriting from this class.
The \textsf{\_\_init\_\_} method receives extra arguments of the conversion
(including other conversions for defining combinators). The \textsf{get\_proof\_term}
function evaluates the conversion on a term $t$ and returns a proof term
showing $t=t'$.

For example, the conversion combinator \textsf{then\_conv} can be implemented
as follows:
\begin{lstlisting}[language=Python]
class then_conv(Conv):
   # Applies cv1, followed by cv2.
   def __init__(self, cv1, cv2):
      self.cv1 = cv1
      self.cv2 = cv2

   def get_proof_term(self, thy, t):
      pt1 = self.cv1.get_proof_term(thy, t)
      pt2 = self.cv2.get_proof_term(thy, pt1.prop.rhs)
      return ProofTerm.transitive(pt1, pt2)
\end{lstlisting}

The concept of tactics can be handled in the same way. In addition, we define
the concept of methods, as general transformations on the current state of
the proof, which directly correspond to actions in the user interface. See \cite{zhan2019}
for more details.

\section{User interfaces for \holpy}\label{sec:user-interface}

The use of a JSON-based theory format means the user interface does not
have to be based on text editing. It also allows different user interfaces
for different domains of application. There are two possible architectures for
user interfaces. In the first architecture, suited for general-purpose theorem
proving, the user interface directly read the JSON file and display it for
the user. Any user changes are directly reflected back to the JSON file.
In the second architecture, suited for theorem proving in special domains,
an intermediate file format is defined for proofs in the special domain. The
user interface reads from the intermediate file format and displays the content
to the user, and reflects changes back to that file format. There is then an
automatic translation process from the intermediate format to the base JSON format.
While the second architecture can also be realized for traditional text-based
proof assistants, the design of \holpy\ makes it more convenient, as the
theory itself can be considered as a Python object.

Currently, we implemented two very different user interfaces based
on the \holpy\ kernel. A point-and-click-based interface for conducting usual
interactive proofs \cite{zhan2019}, and a system for interactive symbolic computation of
definite integrals \cite{li2020}.

The point-and-click-based user interface allows users to carry out a proof largely
by selecting actions from the menu, or from a list of suggestions generated by
the backend. It is designed to be easier to learn than text-based user interfaces.
Note we do not exclude implementing and using a more traditional text-based
user interface in the future. It is possible to envision them operating side-by-side,
depending on the problem and the user's preference.

For interactive computation of definite integrals, we implement an user interface
based on the second architecture. We define an intermediate language recording
computations of definite integrals, with steps such as substitution, integration
by parts, etc. The user interface displays the intermediate format for the user,
and reflects back the changes. At the same time, proof automation is implemented
to translate the format for computation to proofs in higher-order logic. The
proofs are based on a partly formalized library of analysis ported from HOL Light,
built using the point-and-click user interface just described.

Domain-specific user interfaces for interactive theorem proving can be applied to
other areas. One particularly promising direction is program verification. The
goal is to realize an user interface in which the user can edit a program in its own
syntax and insert pre/postconditions and invariants, similar to the current tools
for program verification based on SMT solvers. The resulting verification conditions
can then be proved interactively. When finished, the proofs of individual conditions
are assembled to a complete proof of correctness based on program logic formalized
using a more traditional user interface. We leave implementing these ideas to future work.

\section{Related work}\label{sec:related-work}

Lean \cite{lean} already proposed the use of macros to shorten proof
terms. The virtual machine for Lean evaluates operations on natural numbers,
integers, and arrays directly using C++ functions, rather than within the logic,
resulting in a large speedup for many applications \cite{lean-metaprogramming}.

OpenTheory \cite{opentheory} and Dedukti \cite{dedukti} both provide a
universal language for representing theories in a very general logic,
at least in part with the aim to allow communication between different
proof assistants. Our format for theories bears some similarity to the
formats in these projects. More recently, the projects GamePad \cite{gamepad}
and HOList \cite{holist} built Python interfaces for Coq and HOL Light, respectively,
in order to provide an environment for training deep learning algorithms for automatic
theorem proving. We emphasize that in contrast, our system is self-contained, with
all higher-order logic functions implemented within the system.

Languages for defining proof automation is an active area of
research. Lean's metaprogramming language \cite{lean-metaprogramming}
allows users to define automation in Lean itself. Another research
direction is to add typing to tactic languages
\cite{veriML,mtac}. Other efforts to implement proof assistants in
imperative languages include Lean (C++), Orbital \cite{orbital} (Java) and
the KeYmaera/KeYmaera X tool \cite{keymaera,keymaerax} for reasoning about hybrid systems
(Scala).

\section{Conclusion}\label{sec:conclusion}

In this paper, we reviewed the basic design decisions of the \holpy\ system,
with discussion of the motivation behind these decisions, and the potential advantages they offer.
As illustration, we described two currently implemented user interfaces: one
point-and-click-based interface for general purpose theorem proving, and one specialized
tool for symbolic computation of definite integrals.

This project is still in an early stage, with many improvements planned in the future.
This include automatic reconstruction of proofs from SMT solvers and
resolution-based provers, a foundational treatment of inductive definitions,
better proof automation, and a more complete standard library. In the long term,
we envision further applications in verifying symbolic computation, and applications to
program verification.

\bibliographystyle{splncs04}
\bibliography{paper}

\end{document}